\documentclass[useAMS,usenatbib,fleqn]{mn2e}
\usepackage{listings}
\usepackage{amsmath}
\usepackage{natbib}
\usepackage{graphicx}
\usepackage{color}
\usepackage{rotating}
\usepackage{nicefrac}
\usepackage{txfonts}
\usepackage{trivfloat}
\usepackage{subcaption}

\trivfloat{listfloat}
\usepackage{longtable}
\usepackage[percent]{overpic}
\usepackage{xspace}
\usepackage{url}
\usepackage{mathrsfs}
\usepackage{amssymb}
\usepackage{mathtools} 
\usepackage{enumitem}
\usepackage[percent]{overpic}
\usepackage{soul}

\usepackage{epsfig}
\usepackage{amsmath}
\usepackage{rotating}
\usepackage{natbib}
\usepackage{multirow}
\usepackage[noend]{algpseudocode}

\newcommand{\sima}{\textit{Simulation Archive}\xspace}

\newcommand{\reb}{{\sc \tt REBOUND}\xspace}

\lstdefinestyle{customc}{
  belowcaptionskip=1\baselineskip,
  breaklines=true,
  language=C,
  showstringspaces=false,
  basicstyle=\footnotesize\ttfamily,
}

\usepackage{array}
\usepackage{appendix}
\usepackage{comment}

\def \aap{A\&A}

\def \apjl{ApJ}
\def \apjs{ApJS}

\newcommand{\janus}{{\sc \tt JANUS}\xspace}

\def\gsim{\;\rlap{\lower 2.5pt
 \hbox{$\sim$}}\raise 1.5pt\hbox{$>$}\;}
\def\lsim{\;\rlap{\lower 2.5pt
   \hbox{$\sim$}}\raise 1.5pt\hbox{$<$}\;}

\title{\textsc{JANUS}: A bit-wise reversible integrator for N-body dynamics}
\date{Accepted for publication by MNRAS: \today{}}

\author[Hanno Rein, Daniel Tamayo]{ 
Hanno Rein$^{1,2}$ and Daniel Tamayo$^{1,3,4}$  
\\
$^1$ Department of Physical and Environmental Sciences, University of Toronto at Scarborough, Toronto, Ontario M1C 1A4, Canada\\
$^2$ Department of Astronomy and Astrophysics, University of Toronto, Toronto, Ontario, M5S 3H4, Canada\\
$^3$ Canadian Institute for Theoretical Astrophysics, 60 St. George St, University of Toronto, Toronto, Ontario M5S 3H8, Canada\\
$^4$ Centre for Planetary Sciences Fellow
}

\vspace{0.5\baselineskip}

\begin{document}
\maketitle

\begin{abstract}
Hamiltonian systems such as the gravitational $N$-body problem have time-reversal symmetry.
However, all numerical $N$-body integration schemes, including symplectic ones, respect this property only approximately. 
In this paper, we present the new $N$-body integrator \janus, for which we achieve exact time-reversal symmetry by combining integer and floating point arithmetic.
\janus is explicit, formally symplectic and satisfies Liouville's theorem exactly. 
Its order is even and can be adjusted between two and ten.
We discuss the implementation of \janus and present tests of its accuracy and speed by performing and analyzing long-term integrations of the Solar System.
We show that \janus is fast and accurate enough to tackle a broad class of dynamical problems.
We also discuss the practical and philosophical implications of running exactly time-reversible simulations.

\end{abstract}


\section{Introduction}
\label{sec:intro}
Many astrophysical systems including the gravitational $N$-body problem can be modeled as Hamiltonian systems. 
We approximate all $N$ bodies as point particles with positions $r_i$, velocities $v_i$, and constant masses $m_i$.
Ignoring any external perturbations, the Hamiltonian of the system is then given by
\begin{eqnarray}
    H = \sum_{i} \frac12 m_i v_i^2 - \sum_i \sum_{j<i} \frac{Gm_im_j}{\left|r_i-r_j \right|}. \label{eq:hamiltonian}
\end{eqnarray}
The resulting equations of motion exhibit time-reversal symmetry or time symmetry, i.e. they are invariant under the reversal of the direction of time. 
An equivalent statement is that if we evolve a system for an arbitrary time $t$, reverse the sign of the velocity of each particle, and evolve the system once again for a time $t$, then the system will return to its initial conditions.

In the context of the gravitational dynamics of planetary systems, the equations of motion are too complex to solve analytically except in the simplest cases.
One therefore has to rely on numerical methods to solve for the system's evolution.
Finite floating point representation on a computer introduces roundoff errors that cause deviations from the true trajectory.
Furthermore, integrations forward and then backward in time will not return to the initial point in phase space, since floating point operations are in general irreversible and thus the errors committed in each direction are different.
This is true even for schemes that are {\it formally} time-reversible (i.e., if computers had infinite precision), like leap-frog and many, though not all, implicit and symplectic methods.

It is desirable to have the same symmetries in numerical schemes as in the original dynamical system since they lead to conserved quantities (Noether's theorem).
Furthermore time-reversibility is also intimately related to Liouville's Theorem.
Any integrator that is not bijective (and therefore also not time symmetric) does not keep the phase-space distribution function constant along trajectories, and will introduce numerical dissipation.

In this paper, we present a new formally symplectic integrator \janus that is exactly time-reversible.
We achieve this by replacing critical floating point operations with integer arithmetic.
The force calculation is still subject to round-off errors due to floating point arithmetic.
However, since we make exactly the same errors going forward and backwards in time, the integrator is time-reversible.
\janus is explicit, based on a generalized leap-frog method and its order can be chosen between~2, 4, 6, 8 and~10.
We show that \janus is fast enough to run long-term integration of complex systems like the Solar System over dynamically interesting and astronomically relevant timescales.


The rest of this paper is structured as follows.
In Sec.~\ref{sec:algorithm} we describe bit-wise reversible integrators including the new \janus algorithm. 
Sec.~\ref{sec:implementation} discusses the implementation of \janus.
Then in Sec.~\ref{sec:tests} we present tests of the algorithm, including a long-term integration of the Solar System.
Finally, in Sec.~\ref{sec:discussion} we discuss our results and speculate about implications, both practical and philosophical.

\section{Bit-wise reversible numerical integrators}
\label{sec:algorithm}
We consider an $N$-particle system and assume that the equations of motions can be written in the form 
\begin{eqnarray}
\ddot r_i(t) = F_i(r_0,\ldots,r_{N-1},t) \quad\quad i=0,\ldots N-1.
\end{eqnarray}
Specifically, we assume that the force is not velocity dependent.
This is the case for the equations of motion derived from the Hamiltonian in Eq.~\ref{eq:hamiltonian} and in general from any Hamiltonian that can be split into a kinetic and potential term.

\subsection{Levesque-Verlet integrator}
\cite{LevesqueVerlet1993} presented the first integrator for molecular dynamics that is time-reversible bit by bit. 
The Levesque-Verlet integrator uses the position vectors at times $t_{n-1}$ and $t_n$ to generate new position vectors at time $t_{n+1}$:
\begin{eqnarray} \label{lvi}
    r_i^{n+1} = 2\cdot r_i^{n} - r_i^{n-1} + \left[ h^2 \, F_i^n\right]\label{eq:levesque}
\end{eqnarray}
where $F_i^n$ is the force felt by particle $i$ at time $t_n$ and $h$ is the time step $h=t_{n+1}-t_n=t_n-t_{n-1}$.
With this algorithm, the state of the system is fully specified by the particles' positions at the current and previous timestep, so the particle velocities are never explicitly used or calculated. 
It therefore requires a warmup step at some time $t_0$ to generate the second set of particle positions from the initial conditions.
Aside from this complication, the above algorithm is formally time-reversible. 
To reverse the integration, one only has to swap the two position vectors in Eq.~\ref{lvi},
\begin{eqnarray}
    r^{n+1} &\leftrightarrow& r^{n-1}
\end{eqnarray}
However, a na{\"i}ve implementation would not be reversible bit by bit on a computer.
This is because in floating point arithmetic, the finite representation of numbers leads in general to an irreversible loss of digits under even basic operations like additions and subtractions.
This roadblock can be circumvented by instead storing the positions as integers, whose arithmetic operations are bijective and therefore invertible.
To achieve this, we put a sufficiently fine grid on the computational domain. 
Assuming a typical dynamic range in the coordinates in problems we are trying to solve,
A grid of $2^{64}$ or $2^{128}$ is sufficient to have a relative position accuracy as good or better than that of double precision floating point numbers\footnote{Although this depends somewhat on the range of scales in the problem. For example, a system with a tight binary and an additional wide companion might require a finer grid.}.

The force $F$, however, must still be calculated using standard floating point arithmetic because of square root and division operators. 
These operations cannot be bijectively implemented with integer arithmetic.
After multiplying the force with the timestep squared, we convert (round) from floating point numbers to the nearest integer on our grid. 
We denote this rounding operation with square brackets,~$[\,\cdot\,]$, in Eq.~\ref{eq:levesque}.

Note that even though we perform part of the calculation with floating point operations, the algorithm is nevertheless exactly bit-wise time-reversible. 
The crucial feature is that all the irreversible operations are performed in the middle of the step using the $r^{n}$ coordinates only, so that we re-calculate exactly the same floating point force value (and rounded integer representation) whether stepping forward or backward.
Performing the final operations in integer arithmetic guarantees that the step is exactly time-reversible.

We note that \cite{Rannou1974} and \cite{Earn1992} looked at similar methods to the one described here, but focused on the low dimensional standard map.

\subsection{Leap-frog}
\label{sec:leapfrog}
There are two disadvantages that make the Levesque-Verlet integrator uninteresting for the gravitational N-body problem. 
First, a warmup step needs to be performed at the beginning and end of the simulation, as well as whenever velocity information is needed\footnote{The Levesque-Verlet integrator can be thought of as a kick-drift-kick leap-frog with a specific warmup-step.}.
Second, the integrator is only second order. 
Thus, a relative precision of $\sim10^{-10}$ in a simulation of the Solar System would require a timestep of $\sim0.0008$~days. 
This is not competitive with standard mixed-variable symplectic integrators, which achieve similar precision with timesteps that are $10\,000$ times longer. 

Inspired by the Levesque-Verlet scheme, we came up with a bit-wise reversible version of the standard leap-frog algorithm that is symplectic and can be easily generalized to higher order methods. 

As in the Levesque-Verlet integrator, we use an integer grid to represent the positions of the particles; however, we now also discretize the particle velocities. 
We can then write an integer version of the standard leap-frog integrator:
\begin{eqnarray}
    r_{n+0.5} &=& r_n + \left[ \frac h 2 \; v_n \right]\nonumber\\
    v_{n+1}   &=& v_n + \left[ h \; F_{n+0.5} \right]\label{eq:janus}\\
    r_{n+1}   &=& r_{n+0.5} + \left[\frac h 2 \; v_{n+1} \nonumber\right].
\end{eqnarray}
As before, the force is evaluated using floating point arithmetic, as are the multiplications with the timestep.
We also reuse the square bracket, $[\cdot]$, to denote the rounding operation from floating point numbers to the integer grid.

The nontrivial force calculation that modifies the particle velocities occurs in the middle of the timestep, yielding identical results running forward or backward.
Furthermore, the positions are updated with the velocities evaluated at their respective ends of the timestep. 
These features, combined with the integer grid, yield a formally and bitwise time-reversible algorithm.

Furthermore, the scheme does not require a warmup step, and both velocity and position information is available at every timestep.
To reverse the integration one can either change the sign of the timestep, $h\rightarrow -h$, or apply the involution $(r_n,v_n)\rightarrow (r_n,-v_n)$. 
Note that this works only because the IEEE754 rounding conventions are symmetric about zero.

The first and third step in Eq.~\ref{eq:janus} include a rounding operation because we express the timestep as a floating point number. 
Note that one can also express the timestep as an integer \citep{SyerTremaine1995}. 
If the timestep is an inverse power of two, then the multiplication can be implemented as a simple bit-shift on the integer representation of the velocity.

\subsection{Higher order leap-frog: \janus}
The bit-wise time-reversible leap frog integrator described in Sect.~\ref{sec:leapfrog} can be generalized to higher order with the Baker-Campbell-Hausdorff (BCH) formula \citep[see e.g.][]{Hairer2006}. 
Let us define $\varphi_{\gamma_ih}$ as an operator that evolves the system for a time $\gamma_i\cdot h$ with the bit-wise reversible leap-frog integrator. 
Higher order integrators can then be constructed using compositions of $\varphi$.
For example, consider the operator
\begin{eqnarray}
     \Phi^{(6)}_h = 
    \varphi_{\gamma_1h} \circ 
    \varphi_{\gamma_2h} \circ 
    \varphi_{\gamma_3h} \circ 
    \varphi_{\gamma_4h} \circ 
    \varphi_{\gamma_5h} \circ 
    \varphi_{\gamma_4h} \circ 
    \varphi_{\gamma_3h} \circ 
    \varphi_{\gamma_2h} \circ 
    \varphi_{\gamma_1h}. \label{eq:comp}
\end{eqnarray}
For a suitable choice of constants $\gamma_1,\ldots,\gamma_5$, the operator $\Phi^{(6)}_h$ corresponds to a 6th order integrator in $h$ \citep{Kahan1997,Pihajoki2015}.
Integrators with other orders can be constructed in a similar way.
We have implemented orders 2 (i.e. just the operator $\varphi$),~4, 6,~8, and~10.
We will refer to these integrators as \janus, and more specifically as $\Phi^{(2)}$, $\Phi^{(4)}$, $\Phi^{(6)}$, $\Phi^{(8)}$, and $\Phi^{(10)}$, depending on the order used.
Although it is straightforward to construct even higher order integrators this way, we show in Sect.~\ref{sec:tests} that it seems unlikely that those would be useful for the gravitational $N$-body problem.

We note that for higher order methods, some of the $\gamma_i$ coefficients are negative.
Whereas this gives a formally high order integrator, these methods are in general not very popular. 
In physical terms, the issue is that negative sub-timesteps lead to longer sub-timesteps (in absolute terms) and some of them can become comparable to the original timestep. 
So despite more function evaluations, the integrator does not sample smaller timescales. 
Often a lower order integrator with more timesteps and therefore better sampling of small timescales is more reliable than a high order integrator.
This is the case with any integrator that can be described by Eq.~\ref{eq:comp} and is not specific to bit-wise reversibility.

\begin{figure*}
        \centering
    \begin{subfigure}[t]{0.19\textwidth}
        \includegraphics[width=\textwidth]{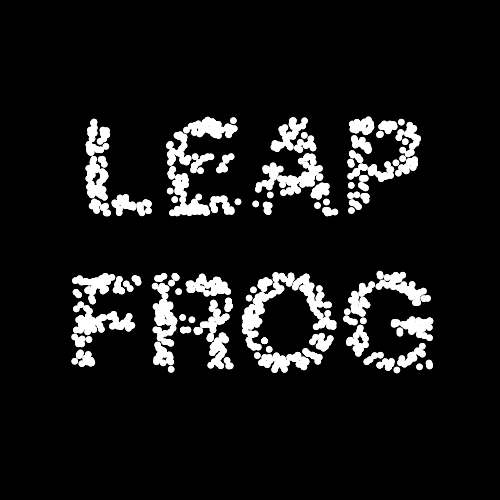}
        \caption{leap-frog, $t=0$}
    \end{subfigure}%
    ~ 
    \begin{subfigure}[t]{0.19\textwidth}
        \includegraphics[width=\textwidth]{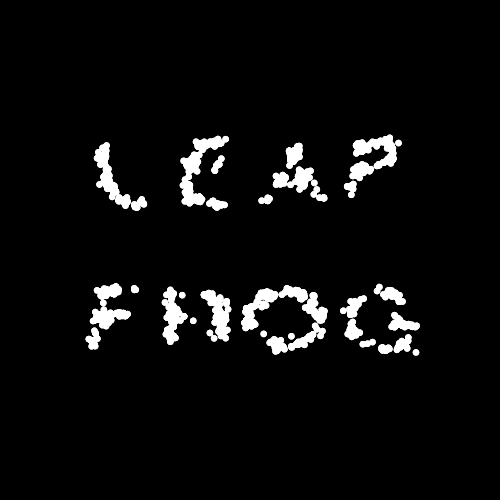}
        \caption{leap-frog, $t=35$}
    \end{subfigure}%
    ~ 
    \begin{subfigure}[t]{0.19\textwidth}
        \includegraphics[width=\textwidth]{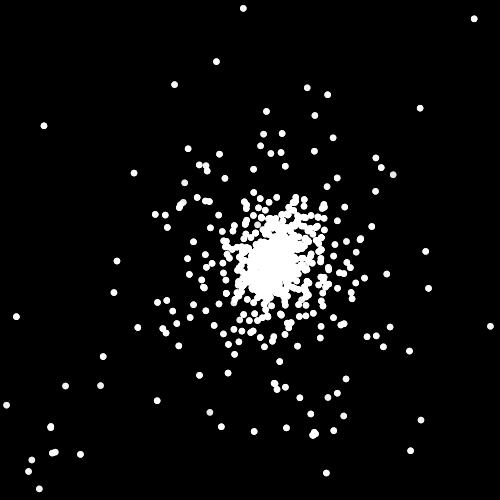}
        \caption{leap-frog, $t=500$}
    \end{subfigure}%
    ~ 
    \begin{subfigure}[t]{0.19\textwidth}
        \includegraphics[width=\textwidth]{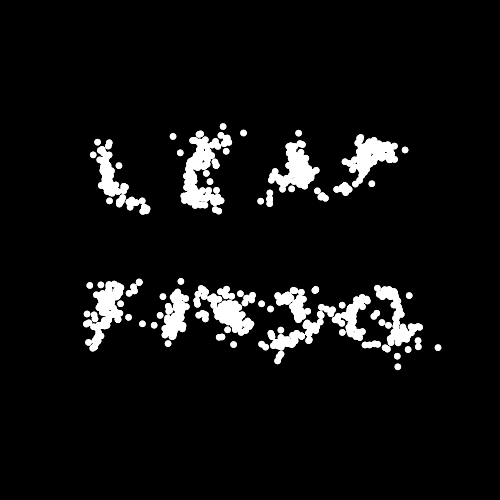}
        \caption{leap-frog, $t=965$}
    \end{subfigure}
    ~ 
    \begin{subfigure}[t]{0.19\textwidth}
        \includegraphics[width=\textwidth]{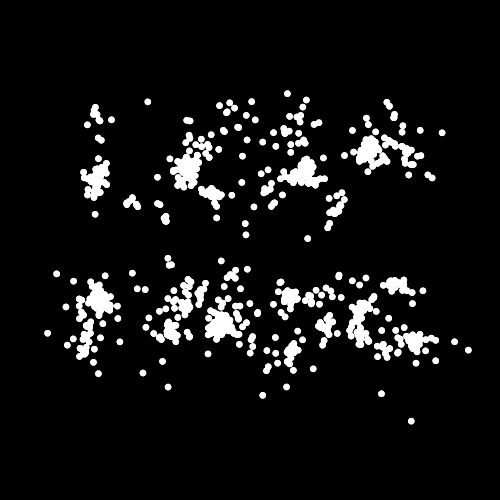}
        \caption{leap-frog, $t=1000$}
    \end{subfigure}
    \vspace{0.6cm}

    \begin{subfigure}[t]{0.19\textwidth}
        \includegraphics[width=\textwidth]{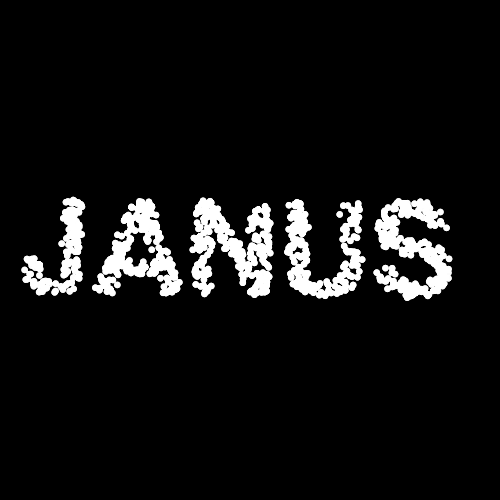}
        \caption{\janus, $t=0$}
    \end{subfigure}%
    ~ 
    \begin{subfigure}[t]{0.19\textwidth}
        \includegraphics[width=\textwidth]{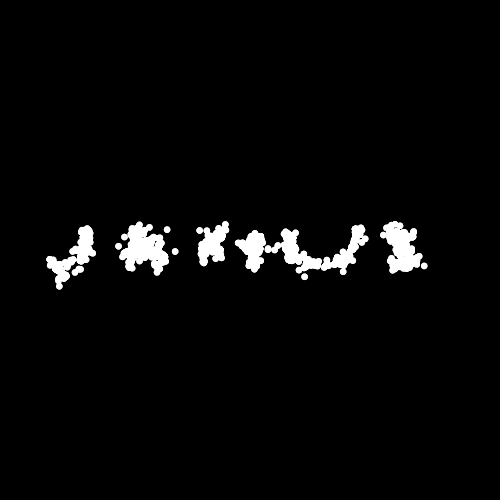}
        \caption{\janus, $t=35$}
    \end{subfigure}%
    ~ 
    \begin{subfigure}[t]{0.19\textwidth}
        \includegraphics[width=\textwidth]{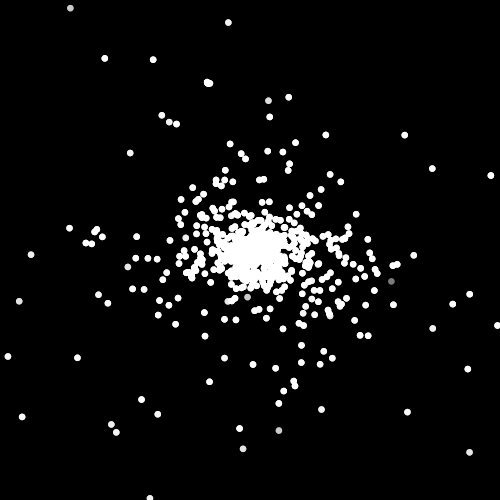}
        \caption{\janus, $t=500$}
    \end{subfigure}%
    ~ 
    \begin{subfigure}[t]{0.19\textwidth}
        \includegraphics[width=\textwidth]{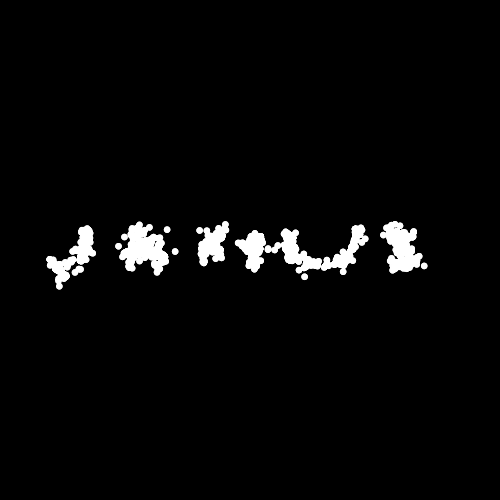}
        \caption{\janus, $t=965$}
    \end{subfigure}
    ~ 
    \begin{subfigure}[t]{0.19\textwidth}
        \includegraphics[width=\textwidth]{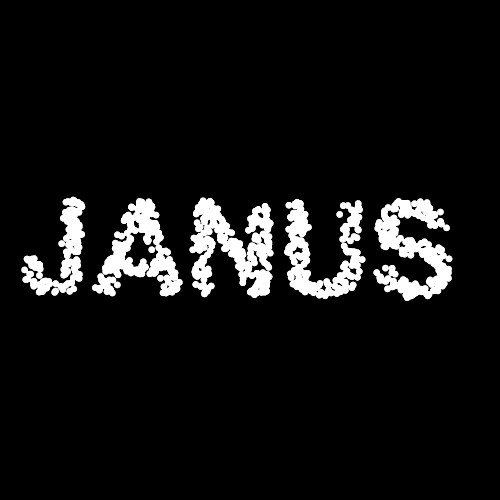}
        \caption{\janus, $t=1000$}
    \end{subfigure}
    \caption{\label{fig:symmetry}Snapshots of two integrations with 1000 gravitationally interacting particles each. The particles are initially at rest in panels (a) and (f). They collapse in the first 500 timesteps due to self-gravity. The sign of the velocities is flipped after 500 timesteps. The system is then integrated for another 500 steps. Because the equations of motion are time symmetric, the initial conditions should be recovered. The first row shows a run with the standard leap-frog integrator which does not reproduce the initial conditions because it is not bit-wise time-reversible. The second row shows a simulation with the new \janus integrator. \janus reproduces the initial conditions exactly because it is bit-wise time-reversible.}
\end{figure*}

\subsection{Symplecticity}
\label{sec:surrogatehamiltonian} 
The question of whether to call \janus symplectic or not is more complicated than it might seem. 

We first note that any numerical $N$-body scheme, including \janus, can only be symplectic to the level of the floating point precision.
An integrator's deviation from perfect symplecticity can be measured directly and it is indeed non-zero for all numerical schemes \citep{Hernandez2016}.
Thus, statements in the literature on whether a given integrator is symplectic refer only to the formal scheme, ignoring the floating point representation of coordinates.
We refer to these integrators as being \textit{formally symplectic}.

If we ignore the floating point errors for a moment, then an important property of such formally symplectic schemes is that they solve a nearby surrogate Hamiltonian exactly\footnote{\textit{Exactly} in the sense that there is no discretization error from the finite timestep.}, explaining their good long-term numerical behaviour \citep[e.g.,][]{SahaTremaine1992}.

A truly symplectic integrator, or in other words a symplectic map, is also a diffeomorphism, i.e. a one-to-one map on the phase-space that is smooth and has a smooth inverse. 
No numerical integrator implemented on a computer can satisfy this condition for two reasons.
First, coordinates always have to be discretized, either on a floating point or integer grid.
Thus the standard notion of smoothness or differentiability breaks down.
Second, all integrators, with the exception of \janus, are also not bijective, a requirement for being a diffeomorphism.  
We could therefore conclude that \janus is \textit{more symplectic} than all other integrators in the sense that it at least is a bijection on the phase-space, even though it is still not a diffeomorphism.

One reason as to why this might matter is that even though \janus is not exactly symplectic in the above sense, it does satisfies Liouville's theorem exactly\footnote{Since we are working on an integer grid, \janus satisfies a discrete version of Liouville's theorem.}.
Liouville's theorem states that the phase-space distribution function along trajectories is constant. 
Satisfying this condition is not equivalent to symplecticity because it does not state anything about the topology.
To our knowledge, \janus is the only $N$-body integrator that does satisfy Liouville's theorem exactly. 
This is a direct consequence of \janus being a bijective operator on the discretized phase-space.

We can show that time reversibility implies bijectivity through a simple proof by contradiction.
Suppose that two phase space points at time $t$ were mapped by the time-reversible \janus scheme onto the same phase space point at time $t_{n+1}$.
Then integrating backwards from the new phase space point by one timestep, one would again arrive at a single point in phase space.
This means that at least one of the trajectories was not time reversible, contradicting our assumption.
Thus time reversibility and bijectivity are equivalent properties of Hamiltonian systems. 
Intuitively, they prevent the contraction or expansion of bundles of trajectories in phase space, as required by Liouville's theorem.

We speculate about the possible implications in the discussion section (Sect.~\ref{sec:discussion}).
For now let us summarize that the family of \janus integrators, $\Phi^{(n)}$, is {explicit, both formally and bit-wise time-reversible, $n$-th order accurate}, exactly satisfying Liouville's theorem, and formally symplectic.

\section{Implementation}
\label{sec:implementation}
We implemented the five bit-wise reversible integrators, $\Phi^{(2)}$, $\Phi^{(4)}$, $\Phi^{(6)}$, $\Phi^{(8)}$, and $\Phi^{(10)}$, which we collectively refer to as \janus, in the open source \reb framework \citep{ReinLiu2012}.
We also modified the \sima \citep{ReinTamayo2017} to work seamlessly with \janus. 

We found that a 64-bit integer grid for both the positions and velocities is well suited for the Solar System case.
Using 32-bit integers would impose a floor on the relative position and velocity accuracy of $2^{-32}\approx 10^{-10}$.
Since the semi-major axes of Mercury and Neptune differ by almost two orders of magnitude, this would translate to relative position and velocity accuracies of at best $10^{-8}$. 
In order to compare results to standard Wisdom-Holman schemes \citep{WisdomHolman1991, Kinoshita1991}, which typically have relative energy errors of less than $10^{-8}$, we need at least 64-bit integers. 
Using 128-bit integers yields no advantages since the forces are evaluated in floating point numbers, which have a relative precision limit of approximately $10^{-16}$.
We thus set 64-bit integers as the default data-type in \janus.
However, our implementation allows us to easily change the integer datatype should another problem require a different precision.

Signed 64-bit integers range from approximately $-10^{19}$ to~$10^{19}$.
To convert from a floating point number $x_f$ to an integer $x_i$, we need to define a grid scale $s$. We can then calculate the integer representation of $x_f$ on our grid $x_i = \left[ x_f/ s\right]$, where the square bracket denotes the rounding operation.
The inverse operation is similar. 
The conversions are applied to the initial conditions and whenever the forces are evaluated.
Typically, one does not have to convert back the positions and velocities to floating point numbers after a timestep, unless an output is required.
We allow the user to specify a separate scale for the positions and velocities.

We note that having positions with 19 digits precision and a force with 16 digits
is a particular sweet spot for running typical long term simulations of the
Solar System and depends on the timestep, the length of the integration,
and the desired final energy error. 
In particular, having more precision in the force calculation does not make a 
difference. On the other hand, having less precision in the force 
calculation would deteriorate the precision of the integrator. In 
practice, we aim to conserve the energy of the system to within 
1 part in 10 billion after one billion orbits. One billion orbits 
corresponds to $10^{13}$ timesteps. Since the error after one timestep 
is roughly 1 part in $10^{16}$ and grows as the square root of the number 
of steps taken (Brouwer's law) the final energy error after 
$10^{10}$ orbits is roughly 1 part in $1^{10}$. 
As we show below, this is consistent with the 
results in Fig.~\ref{fig:janus_longterm}.

\section{Tests}
\label{sec:tests}

\subsection{Time symmetry}
We first test the time-reversal symmetry of \janus. 
To do that we consider a simulation of 1000 gravitationally interacting particles that are initially at rest as shown in panels~(a) and~(f) of Fig.~\ref{fig:symmetry}.
To avoid singularities, the gravitational interactions are softened on small scales.
We integrate this system forward in time for 500 timesteps.
By then the system has collapsed under its own self-gravity, resembling a star cluster.
We then reverse the sign of all velocities and integrate for another 500 timesteps. 
The top row of Fig.~\ref{fig:symmetry} shows snapshots of an integration with a standard leap-frog integrator whereas the bottom row shows snapshots of an integration with our new \janus integrator ($\Phi^{(2)}$).

Both integrators are formally second order, time symmetric, and symplectic. 
One would therefore naively expect to recover the initial conditions at the end of the above procedure.
As one can see in Fig.~\ref{fig:symmetry}, this is not the case for the standard leap-frog integrator.
This is a manifestation of the non-reversibility of floating point operations as described above.

The \janus integrator, on the other hand, recovers the initial conditions.
This can be seen by comparing panels~(f) and~(j) of Fig.~\ref{fig:symmetry}.
The recovery of the initial conditions is exact, down to the last bit of every coordinate of every particle.
We repeated the tests for the higher order version of \janus and for longer timescales.
The results are the identical.

\subsection{Order}
\begin{figure}
 \centering \resizebox{\columnwidth}{!}{\includegraphics{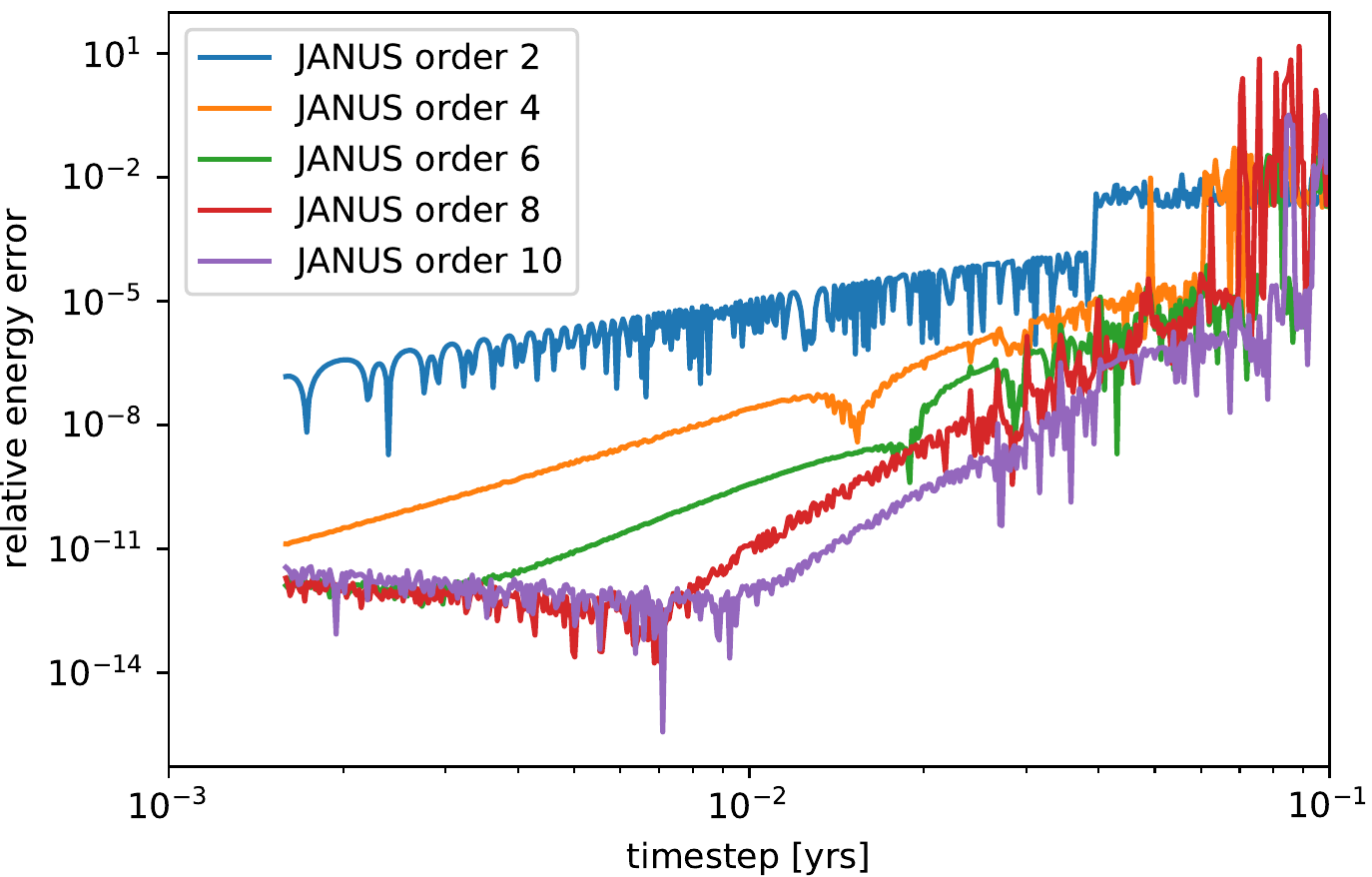}}
 \caption{Relative energy error in simulations of the Solar System with the \janus integrator.
     For large timesteps the error is dominated by the finite timestep. 
     For small timesteps, the floating point errors occurring in force evaluations dominate.
\label{fig:janus_order}}
\end{figure}
Next, we verify the order of \janus.
In this section, and for the rest of this paper, we test \janus on the Solar System, a complex dynamical system that has been studied extensively \citep[for a comprehensive historical review see][]{Laskar2012}.
We use initial conditions from the NASA Horizons System\footnote{\url{http://ssd.jpl.nasa.gov/?horizons}}.
More accurate ephemerides are available \citep{Fienga2011}, but given the other approximations we make (see below), these initial conditions are accurate enough for our purposes.

We integrate all eight planets forward in time for 1000~years with different timesteps.
We set scale parameter for the positions to $10^{-16}$~AU.
Thus, positions given in floating point notation are divided by $10^{-16}$~AU before being rounded to the nearest integer on our grid.
The scale for the velocities is $10^{-16}~2\pi\,\rm{AU}/\rm{yr}$.

The final relative energy error at the end of the integration is shown in Fig.~\ref{fig:janus_order}.
For large timesteps (right side of the plot), the error is dominated by the scheme error, i.e. the finite timestep.
One can see that in this regime (for timesteps greater than $\approx 10^{-2}$), the slope of the curves for integrators $\Phi^{(n)}$ indeed correspond to the order $n$.
For small timesteps (left side of the plot), the error is dominated by the larger number of timesteps required to integrate to a fixed time, and the associated increased accumulation of roundoff errors.
These floating point errors are more important for higher order schemes that evaluate the forces more often than low order schemes
because the high order schemes use more substeps.
We can quantify this somewhat by noting that the 10th order scheme has about twice as many
subteps as the 8th order one. The error scales as the square root of
the number of steps (i.e. approximately a factor of 1.4). Because 
the plot is shown in log-scale over 18 orders of magnitude, these
small factors are hardly visible in the plot. However, one can just
about see that the 10th order curve is higher than the 8th order one.
We note that these errors are due to a computer's inability to calculate the forces exactly, and are unavoidable despite the fact that we have a bit-wise time-reversible integrator.

\subsection{Long term energy conservation}
\begin{figure}
 \centering \resizebox{\columnwidth}{!}{\includegraphics{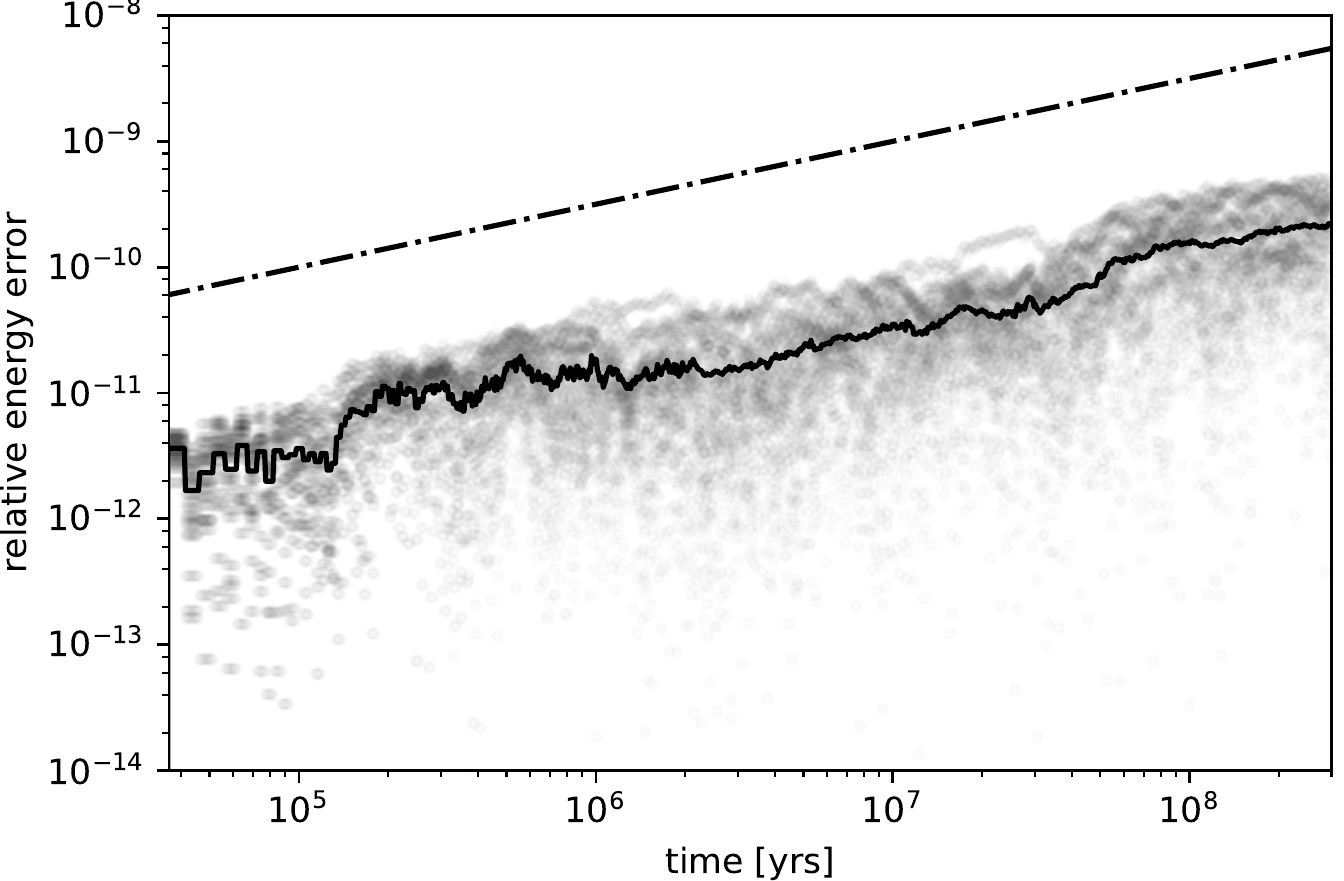}}
 \caption{Relative energy error in long term simulations of the Solar System with the \janus integrator.
Shown are 24 individual simulations in which the initial position of Mercury was perturbed by one meter. 
The black line shows the mean.
The dashed line shows a $\sqrt{t}$ growth, indicating that the integrator is unbiased and following Brouwer's law.
\label{fig:janus_longterm}}
\end{figure}
Let us now test the long-term energy conservation of \janus.
We once again consider the Solar System as a test case.
For each planet, we include a post-Newtonian correction for general relativity in the form of a simple potential \citep{Nobili1986},
\begin{eqnarray}
    \Phi_{\rm GR} &=&  -\frac{3 G^2 m M_\odot^2}{c^2} \frac1{r^2},
\end{eqnarray}
where $m$ is the mass of the planet, $M_\odot$ is the mass of the Sun, $r$ is the heliocentric distance of the planet, and $G$ and $c$ are the gravitational constant and the speed of light, respectively.
This ensures that we reproduce the correct apsidal precession frequencies for the planets, in particular that of Mercury.
We ignore all other non-gravitational effects and all planet-moon systems are treated as single particles.

The Solar System is chaotic with a Lyapunov timescale of approximately 5~Myr \citep{Laskar1989}.
We integrate the system for 60 times longer, i.e. 300~Myr. 
We run 24 different realizations where we perturb the initial position of Mercury by 1~m.
The simulations use a timestep of $0.6$~days and the 6th-order version of \janus, $\Phi^{(6)}$.

The wall-time for a single simulation in our ensemble is 41~days on an Intel Xeon CPU, E5-2697 v2, 2.70~GHz processor.
For comparison, this is about the same performance per timestep as a fast Wisdom-Holman integrator \citep{WisdomHolman1991,ReinTamayo2015}.
However, the timestep in a simulation of the Solar System run with a Wisdom-Holman integrator is typically an order of magnitude larger than what we use here.

The relative energy error as a function of time is shown in Fig.~\ref{fig:janus_longterm}. 
The points show the energy error in 24 individual simulations whereas the solid black line shows the mean.
One can see that during the entire 300~Myr interval, we maintain a relative energy error of $\approx 10^{-10}$.
The dashed line shows an error growth proportional to $\sqrt{t}$.
As one can see, the energy error for \janus follows this sub-linear trend. 
This shows that \janus is unbiased and follows Brouwer's law \citep{Brouwer1937}.

\subsection{Divergence of nearby trajectories and chaos}
\begin{figure}
 \centering \resizebox{\columnwidth}{!}{\includegraphics{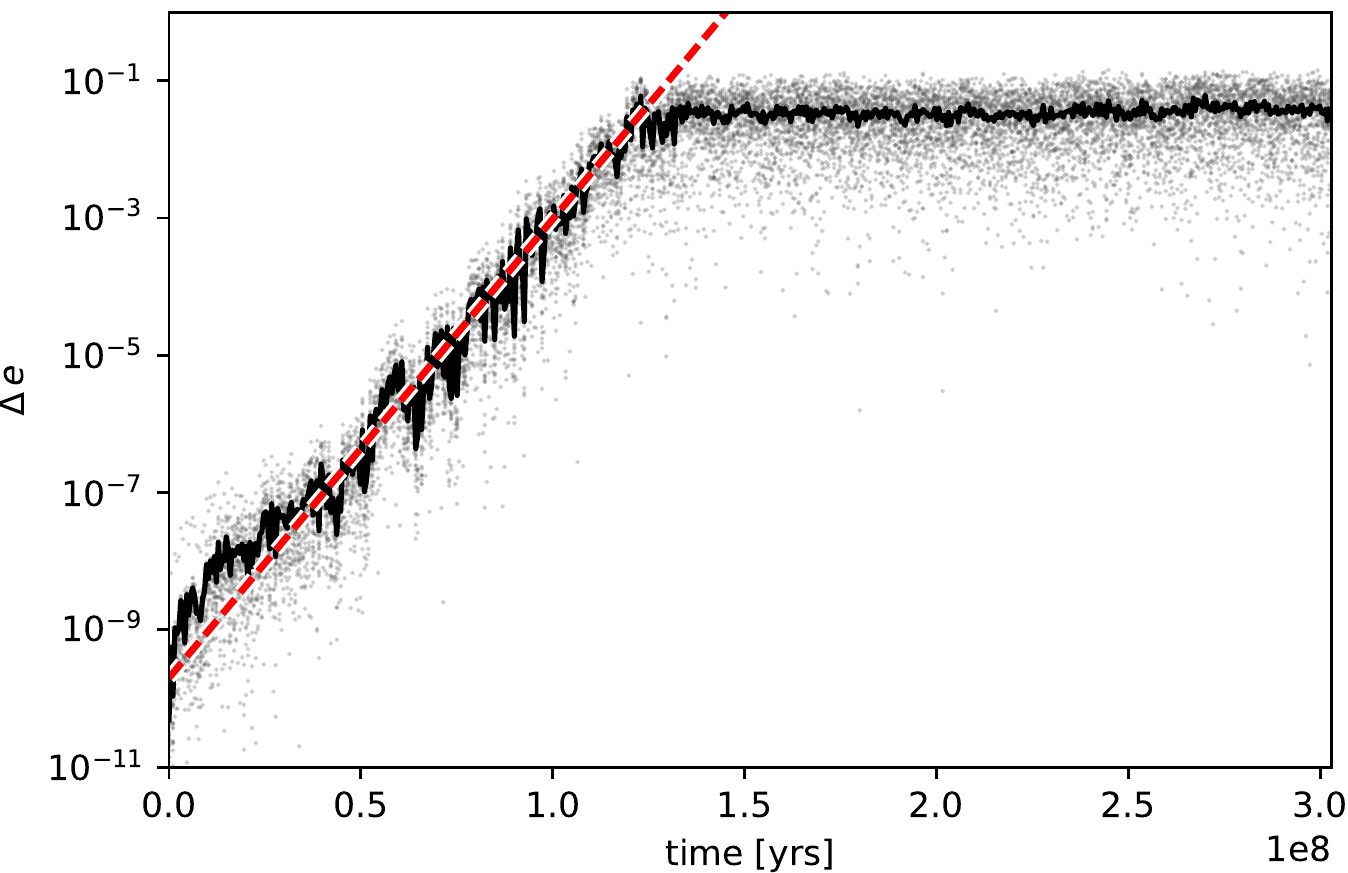}}
 \caption{\label{fig:janus_lyapunov}
 Difference in the averaged eccentricity of Mercury between pairs of simulations of the Solar System.
 The initial position of Mercury was perturbed by $1$~m in each simulation. 
 The gray dots show the difference for two pairs in the ensemble, the solid black line shows the mean. 
 The dashed red line shows an exponential with a timescale of 6.5~Myr, our best fit for the Lyapunov timescale of the inner Solar System.
 }
\end{figure}
The simulations of the Solar System that we presented in the last section will diverge with time because their initial conditions were slightly perturbed. 
There are two reasons for the divergence.
First, because the planets' action variables were perturbed, nearby trajectories will get out of phase.
This effect is typically polynomial with time. 
Second, dynamical chaos will also move initially nearby trajectories further away from each other.
This effect growths exponentially with time.

We can use our simulations to estimate the Lyapunov time, i.e., the timescale for the latter exponential divergence of nearby orbits.
We do this by monitoring slowly varying secular quantities, specifically the eccentricity of Mercury. 
Focusing on variations in these secular quantities (rather than distances in all phase space coordinates) allows us to avoid issues with the quickly growing phase differences.
However, because the eccentricities evolve secularly only approximately\footnote{Small perturbations from nearby planets also change eccentricities on orbital timescales. Often, this is not important. However, it matters for us because we are interested in eccentricity differences of the order of $10^{-9}$. }, we average them over 500\.000~yrs.

The difference in the averaged eccentricity of Mercury between members of our ensemble of simulations is shown in Fig.~\ref{fig:janus_lyapunov}.
The grey dots show the difference for individual pairs.
The black solid curve shows the mean.
The red dashed line corresponds to to an exponential growth with a timescale of $6.5$~Myr.
One can see that any two simulations diverge exponentially fast.
After 120~Myr the differences reach order unity (the eccentricity of Mercury is $\approx 10^{-1}$) and the divergence saturates.
Our measurement of the Lyapunov times in the inner Solar System is consistent with that of previous studies \citep{Laskar1989}.

This test confirms that \janus is able to accurately integrate complex dynamical systems such as the Solar System over extremely long timescales (more than $10^9$ orbital periods of Mercury) and recover chaotic trajectories with high fidelity.

\section{Discussion}
\label{sec:discussion}
In this paper, we presented \janus, the first bit-wise reversible integrator for $N$-body dynamics.
\janus achieves this time-reversal symmetry with a generalized leap-frog scheme that mixes floating point and integer arithmetic.
We implemented different orders of \janus (2, 4, 6, 8 10) which can be chosen by the user at runtime. 

We presented several tests, showcasing that \janus is exactly bit-wise reversible and of the desired order.
We also demonstrated that \janus is capable of accurately integrating complex dynamical systems.
A total of 24 integrations of the Solar System were integrated for $300$~Myr. 
The relative energy error remained below $10^{-10}$, and we were able to recover the chaotic motion of the inner Solar System and measure a Lyapunov time of 6.5~Myr. 

As an illustration of what bit-wise time-reversibility allows us to do, imagine a simulation of the Solar System where Mercury's eccentricity chaotically diffuses to near unity, crossing the orbit of Venus \citep{LaskarGastineau2009}. 
With \janus, we can integrate this realization of the Solar System backwards in time, gradually circularizing the orbit of Mercury, and finally recover the present-day Solar System. 
Similarly, ignoring dissipative effects, we can integrate backwards from the end of a cosmological simulation with bound galaxies at $z=0$ and recapture the adopted primordial density fluctuations\footnote{This should not be confused with an attempt to recover the primordial density fluctuations from the present day matter distribution, which is not possible.}.

It is important to note that the wide variety of robust results using previously available integrators suggests that bit-wise reversibility is not a limiting feature for accurately representing formally time symmetric dynamical systems.
Nevertheless, there are several astrophysical as well as philosophical points and implications that would be fruitful to explore further.

First, and most obviously: if the Hamiltonian system of interest is time symmetric, one would naturally assume that an ideal numerical integrator respects this fundamental property.

Second, to our knowledge \janus is the first N-body integrator that satisfies Liouville's theorem.
The scheme's bijectivity (i.e. reversibility) guarantees that integrated trajectories are unique and thus that phase space distribution functions are constant along trajectories.  
By contrast, the rounding operations in non-reversible schemes cause multiple trajectories to map to the same point in phase space, violating Liouville's theorem. 
In principle, this can lead to unphysical trajectories in phase space. 
Consider a chaotic trajectory that brings the system close to a previously visited point in phase space.
A rounding operation could now shift the system onto the previous point, and incorrectly turn a chaotic trajectory into a periodic one.
From then on, there would be two equally valid histories for the point in phase space, which clearly breaks causality \citep[see Figs.~4 (c) and (d) in][]{Earn1992}.
\janus also suffers from rounding errors, but because the scheme guarantees that mappings in phase space across a timestep are one-to-one, cases like the above are impossible.

Third, imagine a non-periodic realization of the Solar System where a planet is ejected after some time \citep{LaskarGastineau2009}.
By using a bit-wise time symmetric and bijective integrator such as \janus, it is evident that an integration backwards in time must also lead to an ejection of a planet.
Although this might in principle happen only after a \textit{very} long time in the past, we can make this statement without even running the backwards integration using the same argument as above:
Since the orbit is non-periodic, at some point the system will exhaust the volume of phase space that corresponds to bound orbits and a planet will be ejected.
This can be related to Poincare's recurrence theorem \citep{Caratheodory1919}.
  
Fourth, a bit-wise reversible integrator might be useful to construct other integrators. 
For example, it might simplify the development of formally symplectic integrators with adaptive timesteps.  

Fifth, the reversibility properties of \janus could be particularly useful for Hamiltonian Markov Chain Monte Carlo methods. 
There, the time symmetry of the integrator is important to maintain detailed balance \citep{Kendall1994}.

\janus is available within the \reb package which can be downloaded at \url{https://github.com/hannorein/rebound}.
We also provide all notebooks necessary to reproduce the plots in this paper at \url{https://github.com/hannorein/JanusPaper}.
The SimulationArchives \citep{ReinTamayo2017} for the long-term integrations of the Solar System are hosted on zenodo \citep{januszenodo2017}.

\section*{Acknowledgments}
This research has been supported by the NSERC Discovery Grant RGPIN-2014-04553.
We thank David M. Hernandez and Scott Tremaine for helpful discussions in the early stages of this project.
This research was made possible by the open-source projects \texttt{Jupyter} \citep{jupyter}, \texttt{iPython} \citep{ipython}, and \texttt{matplotlib} \citep{matplotlib, matplotlib2}.

\bibliography{full}
\end{document}